\documentclass{elsart}

\usepackage{graphicx}
\usepackage{epsfig}
\usepackage{natbib}
\usepackage{amsmath}
\usepackage[dvips]{color}

\begin{document}
\begin{frontmatter}
\title{Bi-phasic vesicle: instability induced by adsorption of proteins}
\author[add1]{Jean-Marc Allain} and \author[add1]{Martine Ben Amar\thanksref{lbl1}}
\thanks[lbl1]{martine.benamar@lps.ens.fr}

\address[add1]{Laboratoire de Physique Statistique, Ecole Normale 
Sup{\'e}rieure, 24 rue Lhomond, 75231 Paris Cedex 05, France}
\begin{abstract}
The recent discovery of a lateral organization in cell membranes due to small 
structures called 'rafts' has motivated a lot of biological and physico-chemical 
studies. A new experiment on a model system  has shown a spectacular 
budding process with the expulsion of one or two rafts when one introduces proteins on the membrane. In this paper, we give a 
physical interpretation of  the budding of the raft phase. An approach 
based on the energy of the system including the presence of proteins 
is used to derive a shape equation and to study possible instabilities. This 
model shows two different situations which are strongly dependent on the 
nature of the proteins: a regime of easy budding when the proteins are strongly 
coupled to the membrane and a regime of difficult budding.
\end{abstract}
\begin{keyword}
Raft \sep Budding \sep Proteins \sep Membranes \sep Vesicles shape \sep Spherical cap harmonics
\PACS 87.16.Dg \sep 87.10.+e \sep 47.20.Ky
\end{keyword}
\end{frontmatter}

\clearpage
\section{Introduction}

Classical and over-simplified models of  the cell reduces 
the membrane to a bilayer of lipids in a fluid state which is a solvent 
for the proteins of the membrane \cite{Singer72}. But the cell membrane 
is a much more complex and inhomogeneous system. The inhomogeneities
come from a phase separation between small structures called 'rafts' 
\cite{Brown92} and the surrounding liquid phase. These rafts have been 
discovered a decade ago and remain an important issue of cell biology 
but also immunology, virology, etc \cite{Simons97}. A lot of biological 
studies concern the rafts and examine   their composition \cite{Wang03}, 
their in-vivo size \cite{Dietrich02}, their role in signaling \cite{Smith02} or in lipid 
traffic \cite{vanMeer02b} for example. The raft is roughly a mixture of cholesterol 
and sphingolipid but the exact nature of the sphingolipid and its concentration 
can vary between different rafts. In any case and whatever its composition, the 
raft has different physical or chemical properties than the rest of the membrane. 
In this paper, we focus on this specificity which is at the origin of an elastic 
instability that we want to explain.

	Experimentally, the raft in vivo cannot be easily studied and artificial systems 
like GUV (giant unilamellar vesicle) \cite{Dietrich01} appear more appropriate. 
GUV consist in a membrane of lipids with the possibility of a raft inclusion. On 
these artificial systems, a better control of the experimental parameters can be 
obtained and explored. For example, they have been used to study the coupled 
effects of both the membrane composition and  the temperature on the nucleation of 
rafts \cite{Veatch02}. Recently, a new experiment on GUV with rafts has shown a 
spectacular budding process \cite{Staneva03} induced by injection of proteins
called $PLA_2$ (phospholipase $A_2$). 
Before injection, the GUV membrane is in a stable, nearly spherical state.
 But, more precisely,
high-quality  
pictures of vesicles reveal two spherical caps, one for each phase: the raft 
and the fluid phase \cite{webb03}. These two caps have a radius of the same 
order of magnitude (about 5 micrometers, depending on the experimental conditions) 
and are separated by a discontinuous interface. Few seconds after injection of 
$PLA_2$ with a micro-pipette in the vicinity of the raft, one observes a rather strong destabilization 
of the initial shape: the raft tries to rise. The discontinuity of slope at the interface between the 
two caps becomes more and more pronounced. This lifting can be strong enough to 
expel the raft from the vesicle. When two or three rafts are present initially, successive 
expulsions can be observed.	We present here a theoretical treatment showing that the 
driving force of the deformation is the absorption of proteins which locally deforms 
the membrane. We neglect chemical reactions since we focus here on the early 
stages of the instability: the time-scale of the instability is small compared to the 
characteristic time of chemical effects. We restrict ourselves to the simplest model 
relevant for the experiment we want to describe. It  involves standard physical concepts 
of membrane mechanics. The initial shape of the system is given by a minimum of the 
energy of the whole system (that is the inhomogeneous vesicle including proteins). 
A linear  perturbation treatment allows to examine the existence of another solution 
which may lead to a new minimum of energy. This approach is sufficient to predict the 
experimental observation of destabilization and to derive a concentration threshold 
for an elastic instability of the vesicle. The calculation presented 
in this paper concerns only the first stages of the instability. Intermediate stages require 
at least a dynamical  nonlinear calculation including possible chemical effects of 
proteins. The final stage can be achieved by a non-linear calculation or, in case 
of fission, by the energy evaluation of two separated homogeneous spheres 
following the strategy described in \cite{Chen97}.

	Models of vesicles have been widely described in previous papers 
\cite{Helfrich73, Miao91, Seifert91, Jaric95, Julicher96, Dobereiner97}. They 
vary depending on the physical interactions involved taken into account. 
The backbone of all models is based on the minimization of the average curvature 
energy of the bilayer, with the introduction of a possible local membrane asymmetry 
\cite{Canham70, Helfrich73}. A large number of shapes have been predicted in the 
past by this model \cite{Seifert91}. They suitably describe experimental results such 
as the various shapes of red blood cells. Other physical effects can be introduced, 
such as the difference of area between the two layers of the membrane \cite{Seifert97}, 
suggesting a differential compressive stress in the bilayer. These effects are visible under 
 suitable experimental conditions \cite{Dobereiner97}. Here, our scope is to study 
quantitatively the protein-membrane interaction using a generalization of the 
Leibler's model \cite{Leibler86} to an inhomogeneous system. It turns out that this 
model, which describes the proteins as defects on the membrane, leads to a spatially 
inhomogeneous spontaneous curvature which is shown to be responsible for the 
destabilization of the whole system. Going back to the microscopic level, we derive a 
threshold for the protein concentration, which appears as a control parameter. 
Moreover, depending on the shape of the proteins, we are able to select two different 
regimes: a protein-stocking regime and a destabilization regime with possible raft-ejection. 
The idea of a non-homogeneous spontaneous curvature is not new since it has been used for 
mono-phasic vesicles to explain a possible thermal budding \cite{Seifert93}. This does 
not concern the experiment described in \cite{Staneva03} since the temperature is not the relevant control parameter. 
Another scenario for the budding 
process of a raft has been proposed by \cite{Julicher96}: the increase of line tension 
by the proteins leads to an apparent slope discontinuity and to a neck. Again, this 
approach, which is different from ours, is not quantitatively related to the amount of 
proteins. It is why we suggest a different treatment 
 as an interpretation of the raft ejection.

             This paper is organized as follows. Section 2 is devoted to a detailed 
description of the model defining precisely the elastic energy plus the energy of 
interaction combined to the constraints. Section 3 determines an obvious solution 
of the minimization of energy in terms of two joined spherical caps. 
A linear perturbation is performed which gives the threshold of proteins when a 
destabilization occurs. In section 4, the results are analyzed and discussed taking into 
account known or estimated orders of magnitude of physical parameters.

\section{The model}
\subsection{Membrane description}

	The energetic model of the membrane is well established nowadays. 
It can incorporate many different interactions, constraints or restrictions. Here, 
we focus on a precise experiment and we think the model suitable for this 
experiment \cite{Staneva03}. Nevertheless, it can be modified easily for another 
experimental set-up. 

                   We consider a slightly stretched vesicle made of amphiphilic molecules difficult to
solubilize in water. The raft will be denoted by phase $1$, it is usually considered as an  ordered liquid. The remaining 
part is denoted by phase $2$ and is considered as a disordered liquid. As the two phases are liquid, we describe them by two similar 
free energies, each of them having its own set of physical constants. Quantities which 
remain fixed in the experiment are constraints expressed via Lagrange multipliers in the 
free energy. So we define the energy of the bilayer in the phase $(i)$:
\begin{equation}
\label{Fbi}
	F_{b_i} = \int_{S_i}{\left[ { \frac{\kappa _i}{2}H^2 +\kappa_{G_i} K + \Sigma _i} \right]dS}
\end{equation}
with $H$ the mean curvature and $K$ the Gaussian curvature. The square of  $H$ is the classical elastic energy \cite{Helfrich73} 
when we get rid of the spontaneous curvature. Here, there is no physical 
reason to introduce a spontaneous curvature, sign of asymmetry between the two layers. 
The membrane contains enough cholesterol, which has a fast rate of flip-flop and which 
relaxes the constraints inside the bilayer. As for the Gaussian curvature, when a bi-phasic system 
without topological changes is concerned, it gives (Gauss-Bonnet theorem) a mathematical 
contribution only at the boundary \cite{Julicher96}.  
           $\Sigma _i$ means the surface tension: it is the combination between the stretching 
energy of the membrane and an entropic effect due to invisible fluctuations
 \cite{Fournier}. 
In addition to this energy of the bare membrane, we need to introduce the protein-membrane 
interactions.

\subsection{Protein-membrane interactions}	

	Both phases absorb the proteins, as soon as they are  introduced, but probably 
with a different affinity. These proteins are not soluble in water, so we think that they remain 
localized on the membrane and neglect possible exchange with the surrounding bath. As a 
consequence, the number of these molecules remains constant. Moreover, we assume that the 
proteins can not cross the interface \cite{Daumas03}. 
As suggested by S. Leibler \cite{Leibler86}, the average curvature is coupled to the protein 
concentration, for two possible reasons. First, this  can be due to the  conical shape of the proteins 
which locally make a deformation of the membrane. Second, an osmotic pressure on the membrane 
results from the part of the protein in the water \cite{Bickel00}. Whatever the microscopic effect, 
the proteins force the membrane to tilt nearby and thus induce a local curvature. We define the 
energy due to proteins in the phase $(i)$:
\begin{equation}
\label{Fpi}
	F_{p_i} = \int _{S_i}{\left[ { \Lambda _iH\phi+  \lambda _i\phi + \left( { \frac{\alpha_i}{2}(\phi - \phi _{eq_i})^2 
 + \frac{\beta_i}{2}(\nabla\phi)^2 } \right) } \right] dS}
\end{equation}
with $\phi$ the concentration of proteins on the surface. The coupling constant is $\Lambda _i$. 
In Eq.(\ref{Fpi}), $\lambda_i$ is a Lagrange multiplier which allows to maintain the number of 
proteins constant in each phase. The model can be easily changed by considering $\lambda _i$ 
as the chemical potential of the proteins. In this case, the proteins are free to move everywhere 
on the membrane, to cross the interfaces or to go in the surrounding water. The last term in 
Eq.(\ref{Fpi}) is a Landau's expansion of the energy needed to absorb proteins on the surface 
nearby the equilibrium concentration $\phi _{eq}$. The $\phi$ gradient indicates a cost in energy to pay 
for a spatially inhomogeneous concentration. Since the two phases 
are coupled together to  make a unique membrane, let us describe now the interaction between them. 

\subsection{Two phases in interaction}

	The total energy of this inhomogeneous system is the sum of these two individual 
energies for each phase plus at least two coupling terms. First, a more or less sharp interface exists 
between the raft and the phospholipidic part of the membrane. The interface is a line, the cost of 
energy of the transition being given by a line tension $\sigma$ equivalent to the surface tension 
in a vapor-liquid mixing. Second, the surface of the vesicle is lightly porous to the water but not to 
the ions or big molecules present in the solution. So, the membrane is a semi-permeable surface 
and an osmotic pressure appears. As a small variation of the composition of the medium 
surrounding the vesicle induces a large variation of the size of the vesicle, the volume does 
not change when the proteins are injected, however the membrane will break down or transient pores 
will appear \cite{Brochard00}, which is not observed here. 
So, one needs to introduce a Lagrange's 
parameter $-P$ to express the constraint on the volume. Physically, $P$ is the difference of 
osmotic pressure between the two sides of the membrane. Then, the free energy of the bilayer 
becomes:
\begin{equation}
\label{Energie2}
	F_{TOT} = \sum\limits _{i=1,2 } \left( { F_{b_i} + F_{p_i} } \right) + \sigma \int_{ C } {dl} - P\int{dV}
\end{equation}
with $C$ the boundary between the two phases. A variational approach is used to find the initial 
state and to study its stability to   small perturbations.

\section{Static Solution and Stability analysis}

\subsection{Initial state}

\begin{figure}
	\begin{center}
	\includegraphics[width=0.4\textwidth]{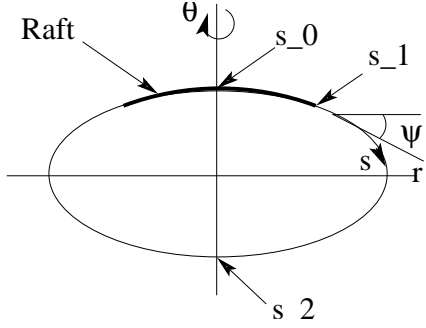}
\caption{Parameterization of an axisymmetric vesicle with two phases in cylindrical coordinates.\label{Param}}
	\end{center}
\end{figure}

	First, we look for the simplest realistic solution with
an homogeneous concentration of proteins in each phase. It
 can be 
found by minimizing the energy~$F_{TOT}$. This minimization gives the Euler-Lagrange equations (E-L equations) 
plus the boundary conditions in an arbitrary set of coordinates. The surface has initially an axis of symmetry. 
Thus, the cylindrical coordinates seem to be the best choice. The parameterization of the surface is 
done by the arc-length $s$ alone (see Fig.\ref{Param}). 
The energy becomes: 
\begin{eqnarray}
\label{energy1}
	F &=& 2\pi \left[ { \int _{s_0} ^{s_1} \mathcal{L}_{1}ds + \int _{s_1} ^{s_2}\mathcal{L}_{2}ds + \sigma r(s_1) } \right] \\
	\lefteqn{\mbox{with} \quad \mathcal{L}_i = \frac{\kappa _i}{2} \left( { \frac{\sin(\psi)^2}{r} + \psi '^2r + 2\psi '\sin(\psi) } \right) 
+ \kappa_{G_i}\sin(\psi)\psi '} \nonumber\\
&&- \Lambda _i\phi \left( {\sin(\psi) + \psi 'r } \right) 
 + \frac{\alpha _i}{2}\phi ^2r + \frac{\beta_i}{2}\phi '^2r + \Sigma '_ir  + \lambda '_i\phi r \nonumber \\
&&- \frac{P}{2}r^2\sin(\psi) 
+ \gamma(r' - \cos(\psi)) \nonumber
\end{eqnarray}
with the new parameters:
\begin{equation}
\label{capi}
 \Sigma _i' = \Sigma _i + \frac{\alpha_i}{2}\phi_{eq_i}^2\quad\mbox{and}\quad
\lambda _i' = \lambda _i - \alpha_i\phi _{eq_i}.
\end{equation}

Minimization with respect to small perturbations of the spatial coordinates ($r$~and~$\psi$) and of 
the protein concentration $\phi$ leads to the E-L equations:
\begin{subequations}
\label{ELg}
\begin{eqnarray}
	\psi '' &=& \frac{\sin(\psi)\cos(\psi)}{r^2} - \frac{Pr}{2\kappa_i}\cos(\psi) - \frac{\psi '}{r}\cos(\psi) + \frac{\Lambda _i}{\kappa _i}\phi ' 
+ \frac{\gamma}{\kappa _ir}\sin(\psi) , \\
	\gamma ' &=& \frac{\kappa _i}{2}\psi '^2 - \frac{\kappa}{2r^2}\sin(\psi)^2 + \Sigma '_i - Pr\sin(\psi) - \Lambda_i\phi\psi ' \nonumber\\
	&&+ \frac{\alpha_i}{2}\phi ^2 + \frac{\beta _i}{2}\phi '^2 + \lambda '_i\phi , \\
	\phi '' &=& -\frac{\Lambda _i}{\beta _ir}(\sin(\psi) + \psi 'r) + \frac{\alpha _i}{\beta _i}\phi - \phi '\frac{\cos(\psi)}{r} + \frac{\lambda '_i}{\beta _i}, \\
	r' &=& \cos(\psi).
\end{eqnarray}
\end{subequations}	
The boundary conditions  deduced from Eq.(\ref{energy1})  at the junction 
  (defined by $s_1$) between the two phases   and the continuity of the radius $r$ give:
\begin{subequations}
\label{CLg}
\begin{eqnarray}
	&&\kappa _{1}\psi '(s_1-\epsilon)r(s_1) + (\kappa _{1}+ \kappa _{G_1})\sin(\psi(s_1-\epsilon)) - \Lambda _{1}\phi(s_1-\epsilon)r(s_1) \\
	\lefteqn{- \kappa _{2}\psi '(s_1+\epsilon)r(s_1) - (\kappa _{2}+\kappa _{G_2})\sin(\psi(s_1+\epsilon)) + \Lambda _{2}\phi(s_1+\epsilon)r(s_1) = 0,} \nonumber\\
	&&\gamma (s_1 - \epsilon) - \gamma (s_1 + \epsilon) + \sigma = 0, \\
	&&\beta _{1}\phi '(s_1 - \epsilon)r(s_1) - \beta _{2}\phi '(s_1 + \epsilon)r(s_1) = 0.
\end{eqnarray}
\end{subequations}

\begin{figure}
	\begin{center}
	\includegraphics[width=0.3\textwidth]{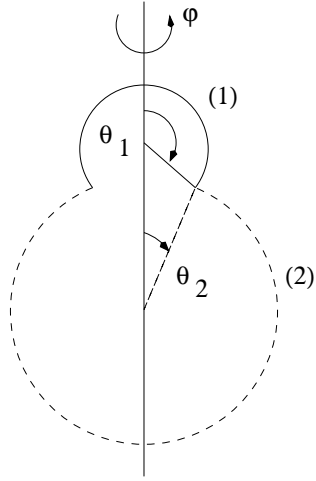}	
\caption{Parameterization of the vesicle near its initial state. The deformation is larger than the real 
one in the initial state.\label{coord}}
	\end{center}
\end{figure}

        The boundary conditions are deduced from the bounds in the variational process. The observation of a shape discontinuity at the boundary between
 the two phases \cite{webb03} suggests a solution which exhibits such a discontinuity 
of the slope at $s=s_1$. 
The angle $\psi$ between the surface and the radius axis is chosen 
discontinuous at the interface (see Fig.\ref{coord}). This allows a tilt of the surface at the boundary $C$. 
The simplest solution, strongly suggested by the experiment, is two spherical caps of radius $R_1$ and $R_2$, one for each phase with a constant concentration $\phi_i$  
(see Fig.\ref{coord}, for clarity, the deformation of the raft is stronger than the real one in the initial state). 
 The minimization shows that one must satisfy the following 
"bulk" conditions for each phase in order to have this solution:
\begin{subequations}
\label{Sphere0}
\begin{eqnarray}
	&&\lambda _i'R_i = 2\Lambda _i - \alpha_iR_i\phi _i \quad \mbox{so}\quad R_i\lambda _i = 2\Lambda _i\\
	&&2\Sigma _i'R_i^2 - PR_i^3 + 2\Lambda _i\phi _iR_i - \alpha_i\phi _i^2R_i^2 = 0.
\end{eqnarray}
\end{subequations}
The boundary conditions give two other relations:
\begin{subequations}
\label{CL0}
\begin{eqnarray}
	\frac{2\kappa_1+\kappa _{G_1}}{R_1} - \Lambda _1\phi _1 =
 \frac{2\kappa_2+\kappa _{G_2}}{R_2} - \Lambda _2\phi _2 ,\\	
	R_2\cos(\theta _2) - R_1\cos( \theta _1) = \frac{2\sigma}{PR_1\sin(\theta _1)}.
\end{eqnarray}
\end{subequations}
where $\theta _1$ and $\theta _2$ are the polar angles in each phase at the boundary (see Fig.\ref{coord}). Note that, for each phase, the couple of equations given by 
Eq.(\ref{Sphere0}) derive the Lagrange parameters like $\lambda_i$ (equivalent to a chemical potential) and $\Sigma_i$  (the tension)
which  are quantities not easy to measure experimentally.
On the contrary, Eq.(\ref{CL0}) give geometrical informations.  
These informations with the other constraints such as the continuity of the radius and the ratio between area of both phases are enough to fix
completely the values of $R_1$, $R_2$, $\theta _1$ and $\theta _2$.
Both conditions have to be satisfied for all protein 
concentrations in order to ensure the existence of the initial homogeneous spherical caps, 
whether they are stable or not. If there is no protein, the conditions (\ref{Sphere0}) reduce to 
$\lambda '=0$ and $2\Sigma _i'R_0^2 = PR_0^3$, which is the classical Laplace equation for 
an interface with a surface tension. Contrary to the law of capillarity, where the surface tension is a physical parameter dependent on the chemical phases involved, the tension here is not a constant characteristic of the lipids of the vesicle. It is a stress (times a length) which varies with the pressure.

\subsection{Linear perturbation analysis}

	Now, we examine  the stability of this solution. Due to the geometry,
 it turns out that the spherical coordinate system is more appropriate here and make the calculations easier.  A perturbation of the spherical cap (i) 
is described by: $R(\theta ; \varphi ) = R_i ( 1 + u(\theta ; \varphi ))$ with $R_i$ the initial radius; in a similar way, a perturbation of the protein 
concentration is $\phi = \phi _i( 1 + v(\theta ;\varphi ) )$ with $\phi _i$ the initial and 
homogeneous concentration of proteins on the surface. We assume that the line tension is 
not modified by the addition of proteins, at least linearly. One can expand the free energy Eq.(\ref{energy1}) 
to second order in $u$ and $v$ in the phase $i$:
\begin{eqnarray}
\label{energy2}
	F_i &=& 2\pi \left[ { \int _{s_{0_i}} ^{s_{1_i}} \mathcal{L}_{i} (u,\nabla u, \Delta u, v, \nabla v)\sin\theta d\theta } \right] 
\quad \mbox{with} \\
	\lefteqn{\mathcal{L}_i = 2\kappa _i \left( { - \Delta u + \frac{1}{4}{\Delta u}^2 + u\Delta u + \frac{{\nabla u}^2}{2} } \right) 
+ \Sigma _i'R_i^2\left( { 2u + u^2 + \frac{{\nabla u}^2}{2} } \right) }  \nonumber \\
&& - 2\Lambda _i\phi _iR_i \left( { u + v - \frac{\Delta u}{2} + \frac{{\nabla u}^2}{2} + uv - \frac{v\Delta u}{2} } \right) - \frac{PR_i^3}{3}(3u+3u^2) \nonumber \\
&&+\frac{\alpha _i}{2}(\phi _iR_i)^2\left( { 2u + 2v + u^2 + 4uv + v^2 + \frac{{\nabla u}^2}{2} } \right) + \frac{\beta _i}{2}\phi _i^2(\nabla v)^2 \nonumber \\
&& + \lambda '_i\phi _iR_i^2 \left( { 2u + v +u^2 + 2uv + \frac{{\nabla u}^2}{2} } \right) \nonumber
\end{eqnarray}
The total free energy is then a function of $u$ and $v$, which allows a variational approach 
to find the Euler-Lagrange's equations (E-L equations) and the  
boundary conditions.
The E-L equations give shapes which are extrema of the free energy. Two sets of 
equations are derived: one for the zeroth order in $u$ and $v$ and one for the first order, 
the energy being calculated up to the second order of the perturbation. The zero-order 
equations gives the same results as Eq.(\ref{CL0}). The first 
order equations  are:
\begin{subequations}
\begin{equation}
\label{EL2a}
	\Lambda _i\phi _iR_i(2u+\Delta u) + \phi _i^2R_i^2\left( {\alpha _iv - \frac{\beta _i}{R_i^2}\Delta v} \right) = 0
\end{equation}
\begin{eqnarray}
\label{EL2b}
	&&\left( { -2\Lambda _i\phi _iR_i + 2\alpha _i\phi _i^2R_i^2 + 2\lambda '_i\phi _iR_i^2 }\right) v + \Lambda _i\phi _iR_i\Delta v \nonumber \\
	&+& \left( { 2\Sigma '_i R_i^2 + \alpha _i\phi _i^2R_i^2 + 2\lambda '_i\phi _iR_i^2 - 2PR_i^3 } \right) u \\
	&+& \left( { 2\kappa _i - \Sigma '_iR_i^2 + 2\Lambda_iR_i\phi _i - \frac{\alpha _i}{2}\phi _i^2R_i^2- \lambda '_i\phi _iR_i^2 } \right) \Delta u 
+ \kappa _i \Delta \Delta u  = 0 \nonumber \\
&&\mbox{equivalent to} \nonumber
\end{eqnarray}
\begin{equation}
\label{EL2c}
        \Lambda _i\phi _iR_i(2v+\Delta v )
         + \left( {  \Lambda _i\phi _iR_i - \frac{PR_i^3}{2} } \right) (2u +\Delta u)+
        \kappa _i \Delta \left( { 2u+ \Delta u} \right)=0
\end{equation}
\end{subequations}

	This coupling imposes boundary conditions which must be treated at the zero-order and the first order. 
We have already studied the zero-order which gives relations (\ref{CL0}). 
As usual for linear perturbation
analysis, the boundary conditions for the perturbation are homogeneous: 
$u=\Delta u=v=0$ at the boundary between
the two phases. Contrary to first intuition and usual procedures, although Eq.(\ref{EL2a}) and (\ref{EL2c}) are linear,
we cannot use the Legendre polynomial basis, due to the specific boundary conditions in this problem. The convenient angular basis in this case turns out to be the spherical cap
harmonics, following standard techniques in geophysics \cite{Haines85} (see Appendix \ref{appendiceA} where we recall some mathematical useful 
relations). These spherical cap harmonics are Legendre functions $P_{x_l}(\cos \theta)$.
 The regular function  at the pole of the cap is of the first kind
and since we restrict on axisymmetric perturbations, these Legendre functions are simply hypergeometric function
$$ {_2}F_1\left( { -x_l,x_l+1,1,\frac{1-\cos\theta}{2} } \right). $$ 
Notice that $x_l$ is not an integer. 
In the case where it is, we recover the Legendre polynomial basis. We select the spherical cap harmonics which vanish
at the boundary angle ($\theta = \theta_1$ or $\theta = \theta_2$, see Fig.\ref{coord}). This condition at the boundary gives
a discrete infinite set of non-integer $x_l^{(i)}$ values for the phase ($i$). 
$l$ is an integer index used to order the allowed values $x_l^{(i)}$ by increasing values. It is also the number of
zero of the function $P_{x_l^{(i)}}$ on the cap. These harmonics have the properties 
to be an orthogonal basis and to be eigenfunctions of the Legendre equation with eigenvalues: $x_l^{(i)}(x_l^{(i)}+1)$. 
	We define $u(\theta)=\Sigma_l u_l P_{x_l}(\cos \theta)$ and
$v(\theta)=\Sigma_l v_l P_{x_l}(\cos \theta)$. From the first E-L equation 
(\ref{EL2a}), one can deduce the amplitude
 $v_{i,l}$ 
 the protein concentration from $u_{i,l}$ in the phase ($i$):
\begin{equation}
\label{amplitude}
	v_{i,l} = \frac{\Lambda _i \left[ { x_l^{(i)}(x_l^{(i)}+1)-2 } \right]}{\phi_iR_i\left[ { \alpha_i+\beta_i(x_l^{(i)}+1)x_l^{(i)}/R_i^2 } \right] }u_{i,l}.
\end{equation}
We  introduce  $q^2 = x_l^{(i)}(x_l^{(i)}+1)/R_i^2$, which is similar
 to the spatial period 
of the perturbation. Then, from the second E-L equation (\ref{EL2b}) and after elimination using Eq.(\ref{amplitude}), we derive 
\begin{equation}
\label{seuil}
	\Lambda _i^2(q^2-2/R_i^2) = \left( { \Sigma _i' -\frac{\alpha_i}{2}\phi _i^2 + \kappa _i q^2 } \right)(\alpha_i +\beta_iq^2).
\end{equation}

        Our result can be compared to previous analysis made
 in two different asymptotic limits in the homogeneous case. 
In these cases, the cap is a complete sphere and $x_l$ is an integer.
First, for $\beta=0$, we recover the result of \cite{Mori93} for an
homogeneous vesicle without diffusion. Second, when $R_i$  goes to
infinity, we recover the result for an homogeneous flat membrane \cite{Leibler86}. 

	Notice that, in Eq.(\ref{seuil}), the protein concentration has 
a similar significance as a negative 
 surface tension: one can make the change of variable
$\Sigma _i'' = \Sigma _i' - \alpha_i\phi^2 _i/2=
\Sigma _i + \alpha_i(\phi_{eq_i}^2-\phi^2 _i)/2$. The principal effect of
the proteins is to decrease the surface tension which is an obvious sign of instability.

\section{Discussion}

We will use the protein concentration $\phi _i$ as our control parameter. 
Eq.(\ref{seuil}) gives for each mode $x_l^{(i)}$ a threshold concentration
$\Phi _i$ such that for $\phi _i\le \Phi_i$ the initial state is stable and for $\phi _i\ge\Phi_i$,
one of the two phases is unstable, leading to a complete instability. The threshold concentration 
$\Phi _i$ strongly  depends on the physical properties of each phase. Then, the
two parts of the initial system have no reason to be unstable simultaneously.
The deformation of the other phase (not unstable to linear order) will be induced 
by the non-linear effects not included in this analysis.

       Since the thermal energy $kT$ is the only external energy and 
the typical length of the phase (i) is $R_i$, one can introduce dimensionless 
parameters: $\tilde{q} = x_l(x_l+1)$, ${\kappa}_i = \kappa _i/kT$, 
$\tilde{\Sigma}_i' = \Sigma _iR_i^2/kT$, $\tilde{\Lambda} _i = \Lambda_i/(kTR_i)$,
$\tilde{\alpha}_i = \alpha_i/(R_i^2kT)$ and $\tilde{\beta}_i =\beta_i/(R_i^2L_c^2kT)$ 
with $L_c$ a characteristic length for the gradient of protein concentration. Then, 
the protein concentration is replaced by $R_i^4\phi _i^2$ which is proportional to 
the square of the number of proteins in the phase (i).

Rewriting the threshold (\ref{seuil}), we find for the threshold concentration, in dimensionless parameters: 
\begin{equation}
\label{phi_l2}
	R_i^4\Phi _i^2 = \frac{2}{\tilde{\alpha}_i}\left( { \tilde{\Sigma}_i'+ 
	\tilde{\kappa}_i\tilde{q}^2-\frac{\tilde{\Lambda}_i^2(\tilde{q}^2-2)}
	{\tilde{\alpha_i}+\tilde{\beta}_i(L_c/R_i)^2\tilde{q}^2} } \right).
\end{equation}	
From Eq.(\ref{phi_l2}), the search of the smallest threshold  concentration 
gives  two different regimes depending on the value 
of the dimensionless constant:
\begin{equation}
	c = \frac{\tilde{\Lambda _i}^2\left( { \tilde{\alpha_i}+2\tilde{\beta_i}\frac{L_c^2}{R_i^2} }\right) }{\tilde{\kappa_i}\tilde{\alpha_i}^2}
 = \frac{\Lambda_i^2(\alpha_i+2\beta_i/R_i^2)}{\kappa_i\alpha_i^2}.
\end{equation}
This constant describes the strength of the coupling of the protein with the membrane  ($\Lambda _i$) to the resistance 
of the membrane ($\kappa _i$) and to the absorption power ($\alpha _i$).
\begin{figure}
	\begin{center}
	\includegraphics[width=0.5\textwidth]{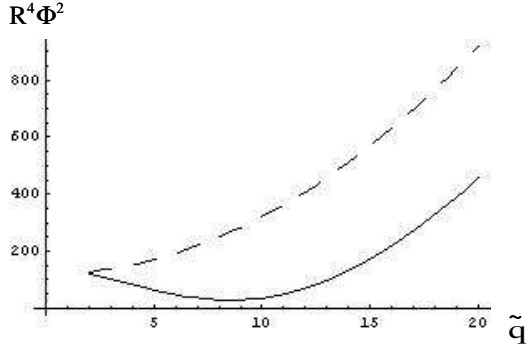}
	\caption{Threshold concentration versus the reduced mode$\tilde{q}$ for two 
different values of (\ref{phi_l2}) are taken as $\tilde{\alpha}_i = 1$, $\tilde{\kappa}_i = 1$, 
$\tilde{\Sigma}_i' = 60$, $\tilde{\beta}_i(L_c/R_i)^2 = 0.01$. The solid line is the case of 
large tilt due to each protein: $\tilde{\Lambda}_i = 1.7$. The dashed curve is the case of 
a small tilt: $\tilde{\Lambda}_i = 0.01$.\label{courbe}}
	\end{center}
\end{figure}
In the weak interaction regime ($c \le 1$), the protein concentration is an increasing 
function of $\tilde q$ (see fig.~\ref{courbe}). 
So, the threshold is obtained for the smallest possible $x_l$: $x_0$. The direction of
the deformation (inside or outside the initial cap) would be deduced from a third order calculation or 
from a numerical simulation. According to the definition of $\Sigma _i'$
(Eq.(\ref{capi})), the concentration 
required to destabilize the membrane is found bigger than the equilibrium concentration. 
But in this case, one expects that this threshold $\Phi _i$ is difficult or impossible to reach 
since it requests the absorption of a concentration of proteins larger than $\phi_{eq_i}$: 
probably, the excess of proteins would  prefer to dissolve in the surrounding water then 
forming  aggregates. So, the weak regime of instability is not observable
 experimentally, our basic state made of two spherical caps is stable and
proteins are stocked only.

	In the strong interaction regime, the limiting concentration shows a minimum not necessary 
for the smallest $x_l$ 
(see fig.\ref{courbe}). This has two consequences. First, the limiting concentration is less 
than the equilibrium concentration and it is easier to induce the instability. Second, the 
first unstable mode could be modified: $\tilde{q}\approx 10$ for the chosen numerical values. So, we have 
something more complex than the simple oblate/prolate ($x_0$) deformation. This regime is observable 
and probably corresponds to the observed  shape instability. 
In any case, our basic state cannot be seen in the experiment except as a transient.

	The existence of these two regimes, depending on the nature of the proteins, allows 
two possible and distinct scenarios for the cell: there is no doubt that
 this property is useful and probably used  for   
biological purpose. The  main difficulty of  this study is the quantitative determination of the parameters 
since experimental values are not available even for this minimal model. Let us estimate $c$. 
The curvature of the membrane is of order $1/R_i$, its surface is close to $R_i^2$. $\kappa$ 
is of order $10kT$ for an unstretched vesicle, so $\tilde{\kappa} \approx 10$. $\alpha$, the cost 
of energy needed to increase the concentration of proteins, can be deduced from the 
energy required to remove a protein from the surface which is is about $100kT$. It changes 
the concentration of proteins of $1/R_i^2$ so $100kT = \alpha_i/R_i^2$ and $\tilde{\alpha}\approx 100$. 
$\beta$ is deduced from $L_c$ which should be a small fraction of the radius of the sphere. 
We will take hereafter $L_c \approx R_i/10$. The proteins are moving at the surface of the 
membrane due to the Brownian motion. So the energy to move one protein by the length 
$L_c$ is about $kT$. Then, $\beta _i =kTL_c^2R_i^2$ and $\tilde{\beta _i} \approx 1$. The value 
of $\Lambda _i$ is more difficult to determine. $\Lambda _i$ is the coupling constant 
between the membrane and the proteins. This is expressed by the spontaneous curvature 
radius $R_P$. If $R_P$ is small, the coupling effect is strong and, on the opposite, if $R_P$ 
is large, $\Lambda _i$ is small, which suggests that $\Lambda _i$ is proportional to $1/R_P$. 
But $\Lambda _i$ is an energy multiplied by a length. Then, a good order of magnitude for 
$\Lambda _i$ is $kTR_i^2/R_P$, so $\tilde{\Lambda _i} = R_i/R_P$. Finally, we get 
$c \approx R_P/R_i$. If $R_P$ is smaller than $R_i$, the system is in the strong coupling regime 
and in the other case, the interaction between membrane and proteins is weak. When the two 
phases have approximately the same physical constants \cite{webb03}, the instability occurs first in the 
largest phase, as shown by Eq.(\ref{phi_l2}) in the previous conditions. 
Nevertheless, the ejection of one part of the membrane requires a complete 
nonlinear dynamical treatment which will be derived from this energy formulation.

\section{Conclusions}
        We have proposed a model of instability for an
inhomogeneous vesicle which absorbs proteins. This instability   is at the origin of a separation into two
vesicles, one for each phase
 as seen experimentally.  Our model rests on
a "bulk" effect and
assumes that the proteins are distributed everywhere on the
membrane contrary to the
"line tension "model which assumes a high concentration of the
proteins at the raft
boundary. To validate (or invalidate) our model, an experimental
test could be the use of
phosphorescent proteins with the same properties. It would be a way
to follow the place where the
proteins prefer to diffuse and stay on the membrane. Although we
ignore the feasibility of
such an  experiment, it would provide a very useful information.

We thank G. Staneva, M. Angelova and K. Koumanov   for
communicating their results prior to publication. We acknowledge
enlightening discussions with J.B. Fournier.


\appendix
\section{Spherical cap harmonics}
\label{appendiceA}

	The spherical cap harmonics are eigenvalues of the Laplace's equation in spherical coordinates. The Laplace's problem 
can then be rewritten as a Legendre's equation. The general solution is:
$$U^m_n = f(\phi)L^m_{x_l}(\cos\theta)$$
with $\theta$ the colatitude, $\phi$ the longitude and $L^m_{x_l}$ an associated Legendre function. 
The eigenvalues associated to this solution are $m^2$ and $-x_l(x_l+1)$. So the solutions are symmetric with respect to $x_l = -1/2$.
So we can restrict to $x_l \ge -1/2$ in all cases.

	Generally speaking, $m$ and $x_l$ can be integer, real or even complex and are determined by the boundary conditions. 
For a sphere, the solution must be periodic in the $\phi$ angle. This implies $m$ real. 
In the particular case of an axisymmetric solution, which is the case in this paper, $m = 0$.

	The boundary condition on $\theta$ for $\theta = 0$ is a condition of regularity:
\begin{eqnarray}
\label{theta0}
	\frac{\partial U^0_{x_l}}{\partial \theta} = 0 \quad \mbox{for} \quad m=0 \\
	\nonumber \\
	U^m_{x_l} = 0 \quad \mbox{for} \quad m\neq 0
\end{eqnarray}
It is satisfied by the Legendre functions of the first kind and excludes those of the second kind. Notice that this condition is required both for
a complete sphere and for a spherical cap. 
	
	In the case of the sphere, the boundary condition $\theta = \pi$ is similar to Eq.(\ref{theta0}). 
The values of $x_l$ are then integer and the solutions are the classical associated Legendre polynomials.

	For a spherical cap whose ends are given by $\theta = \pm \theta _0$, the boundary conditions at $\theta_0$ are given by standard physical requirements. These boundary conditions can be 
satisfied by using two kinds of solutions such that either:
\begin{equation}
\label{cond1}
	\frac{\partial U^m_{x_l}}{\partial \theta} = 0 \quad \mbox{for} \quad \theta = \pm\theta _0
\end{equation} 
or:
\begin{equation}
\label{cond2}
	U^m_{x_l} =0 \quad \mbox{for} \quad \theta = \pm\theta _0
\end{equation}
These conditions are satisfied by Legendre functions $P^m_{x_l}(\cos\theta)$ with $x_l$ not necessary integer. No function can 
satisfy simultaneously the conditions (\ref{cond1}) and (\ref{cond2}) and there is two sets of $x_l$ which depend on the $m$ value. We call
$y_l(m)$ the values of $x_l$ such as (\ref{cond1}) is satisfied and $z_l(m)$ the values of $x_l$ such as (\ref{cond2}) is satisfied.

	Functions in one set are orthogonal to each other but are not orthogonal to those of the other set. It is easy to show that:
\begin{eqnarray}
	\lefteqn{\int _0^{\theta _0} {P^m_{y_{l_1}(m)}(\cos\theta)P^m_{y_{l_2}(m)}(\cos\theta)\sin\theta d\theta} = 0 \quad \mbox{for}\quad l_1\neq l_2} \nonumber\\
	\lefteqn{\int _0^{\theta _0} {P^m_{z_{l_1}(m)}(\cos\theta)P^m_{z_{l_2}(m)}(\cos\theta)\sin\theta d\theta} = 0 \quad \mbox{for}\quad l_1\neq l_2 }\nonumber\\
	\lefteqn{\int _0^{\theta _0} {P^m_{y_{l_1}(m)}(\cos\theta)P^m_{z_{l_2}(m)}(\cos\theta)\sin\theta d\theta} =} \nonumber \\
	&&\quad\quad\quad -\frac{\sin\theta _0P^m_{y_{l_1}(m)}(\cos\theta)\{[P^m_{z_{l_2}(m)}(\cos\theta)]/d\theta\}}{(y_{l_1}-z_{l_2})(y_{l_1}+z_{l_2}+1)} \nonumber
\end{eqnarray}
	
\begin{figure}[h!]
	\begin{center}
	\includegraphics[width=0.5\textwidth]{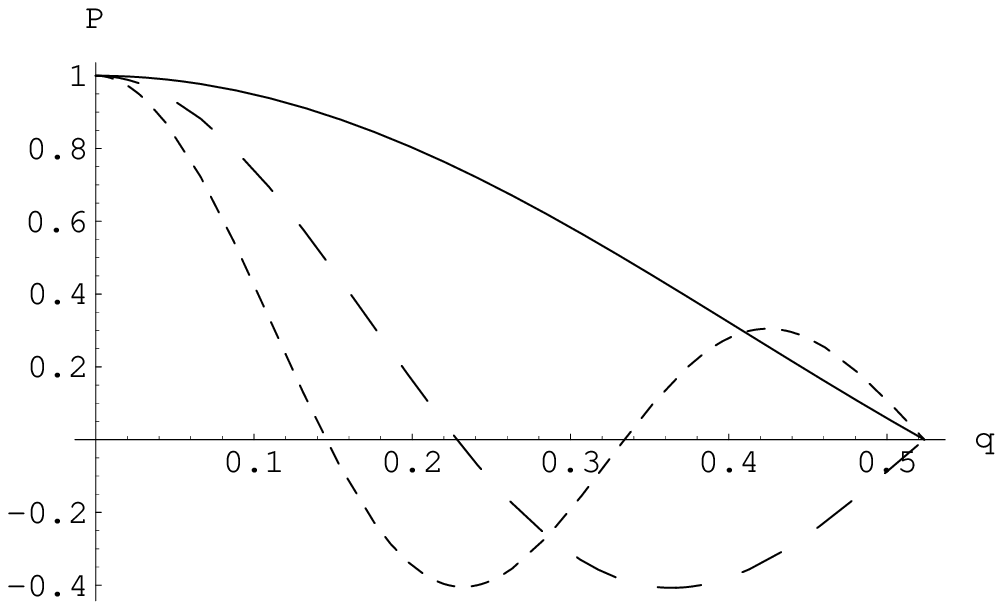}
	\caption{Legendre functions $P_{y_l}$ versus the angle $\theta$ for $\theta _0 = \pi /6$ and for $l=0$ (solid curve), $l=1$ (large dashing) and $l=2$ (small dashing)\label{L1}}
	\end{center}
\end{figure}

\begin{figure}[h!]
	\begin{center}
	\includegraphics[width=0.5\textwidth]{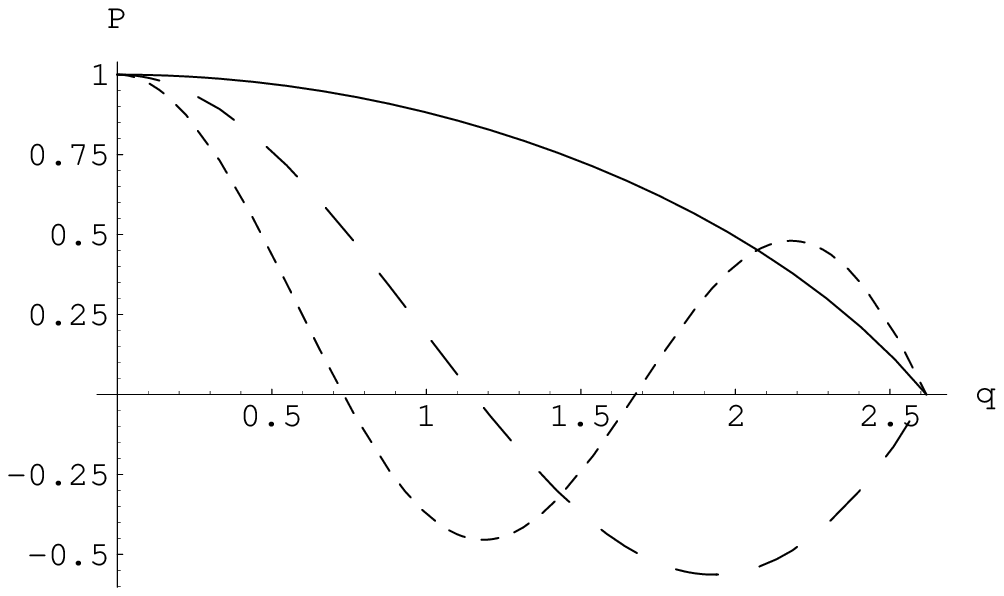}
	\caption{Legendre functions $P_{y_l}$ versus the angle $\theta$ for $\theta _0 = 5\pi /6$ and for $l=0$ (solid curve), $l=1$ (large dashing) and $l=2$ (small dashing)\label{L2}}
	\end{center}
\end{figure}

\begin{figure}[h!]
	\begin{center}
	\includegraphics[width=0.7\textwidth]{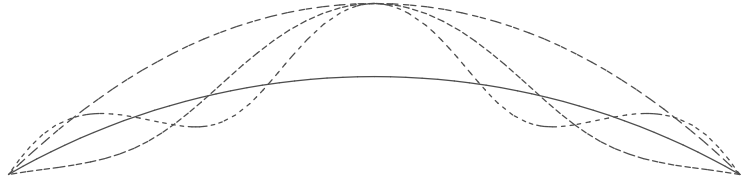}
	\caption{Deformations of a spherical cap of radius 1 and half-angle $\pi/6$. The deformations are dues to Legendre functions $P_{y_l}$. For clarity, the maximum amplitude of the deformation is fixed to 0.1. This value is too large in principle for the linear analysis. The solid curve is the initial state. The largest dashing is the case $l=0$. The intermediate dashing is the case $l=1$. The smallest dashing is the case $l=2$.\label{L3}}
	\end{center}
\end{figure}

	If the physics requires the boundary condition (\ref{cond1}) or (\ref{cond2}), the set of solutions $y_l$ or $z_l$ is enough to form a 
basis of solution of the problem. In the other case, one have to combine both of them and the resolution of the complete problem becomes more harder.
In the case of this paper, we focus on the case of axisymmetric solutions $(m=0)$. The boundary conditions are given by \ref{cond2}, so the good set of parameters are the $y_l(0)$. The table \ref{table1} presents the first values of $y_l$, calculated for two angles $\theta_0$ ($\pi /6$ and $5\pi /6$), chosen as example. The figure \ref{L1} shows the three first $P_{s_l}$ for $\theta _0 = \pi /6$. The figure \ref{L2} shows the three first $P_{x_l}$ for $\theta _0 = 5\pi /6$. The figure \ref{L3} shows the deformation of the spherical cap in the case of a perturbation by the three first $P_{x_l}$ for $\theta _0 = \pi /6$.

\begin{table}[h!]
\begin{center}
\begin{tabular}{c|c|c|c|c|c|c|c}
	$l$ & 0 & 1 & 2 & 3 & 4 & 5 & 6 \\ \hline
	$\theta _0 = \pi /6$   & 4.08 & 10.04 & 16.03 & 22.02 & 28.01 & 34.01 & 40.01 \\ \hline 
	$\theta _0 = 5\pi /6$ & 0.35 & 1.57 & 2.78 & 3.98 & 5.19 & 6.39 &7.59 \\
\end{tabular}
\caption{Values of $y_l$ for $m = 0$ for the $7$ first values of $l$.\label{table1}}
\end{center}
\end{table}

\pagebreak

\end{document}